\documentclass[aps,twocolumn,preprintnumbers,amsmath,amssymb,nofootinbib,superscriptaddress,notitlepage]{revtex4}
\usepackage{graphicx,color,dcolumn,booktabs,bm}
\usepackage{longtable,lscape}
\usepackage{txfonts}
\usepackage{overpic}
\usepackage{amssymb}
\usepackage{gensymb}
\usepackage{indentfirst}
\usepackage{feynmf}
\usepackage{slashed}
\usepackage{cases}
\usepackage{multirow}
\usepackage{appendix}
\usepackage[figuresright]{rotating}
\usepackage[colorlinks,
citecolor=blue,
anchorcolor=red,
menucolor=red,
linkcolor=red,
filecolor=red,
runcolor=red,
urlcolor=blue,
frenchlinks=red]{hyperref}
\usepackage{epstopdf}
\usepackage{bm}
\usepackage{tabularx}
\usepackage{threeparttable}
\usepackage{bbding}
\usepackage{array}
\usepackage[section]{placeins}
\usepackage{makecell}
\usepackage{comment}
\makeatletter

\newcommand{\Rmnum}[1]{\expandafter\@slowromancap\romannumeral #1@}
\makeatother

\usepackage{soul}

\begin{document}
	\title{Non-$D\bar{D}$ decays into light meson pairs of the $D$-wave charmonium $\psi_3(3842)$}
	\author{Zi-Yue Bai}
	\email{baizy15@lzu.edu.cn}
	\author{Bao-Jun Lai}
	\email{laibj2024@lzu.edu.cn}
 
	\affiliation{School of Physical Science and Technology, Lanzhou University, Lanzhou 730000, China}
	\affiliation{Lanzhou Center for Theoretical Physics, Key Laboratory of Theoretical Physics of Gansu Province, Lanzhou University, Lanzhou 730000, China}
	\affiliation{Key Laboratory of Quantum Theory and Applications of MoE, Lanzhou University, Lanzhou 730000, China}
	\affiliation{Research Center for Hadron and CSR Physics, Lanzhou University and Institute of Modern Physics of CAS, Lanzhou 730000, China}
	\author{Qin-Song Zhou}
	\email{zhouqs@imu.edu.cn}
	\affiliation{Lanzhou Center for Theoretical Physics, Key Laboratory of Theoretical Physics of Gansu Province, Lanzhou University, Lanzhou 730000, China}
	\affiliation{Research Center for Hadron and CSR Physics, Lanzhou University and Institute of Modern Physics of CAS, Lanzhou 730000, China}
	\affiliation{MoE Frontiers Science Center for Rare Isotopes, Lanzhou University, Lanzhou 730000, China}
	\affiliation{Center for Quantum Physics and Technologies, School of Physical Science and Technology, Inner Mongolia University, Hohhot 010021, China}
	\author{Xiang~Liu\footnote{Corresponding author}}
	\email{xiangliu@lzu.edu.cn}
	\affiliation{School of Physical Science and Technology, Lanzhou University, Lanzhou 730000, China}
	\affiliation{Lanzhou Center for Theoretical Physics, Key Laboratory of Theoretical Physics of Gansu Province, Lanzhou University, Lanzhou 730000, China}
	\affiliation{Key Laboratory of Quantum Theory and Applications of MoE, Lanzhou University, Lanzhou 730000, China}
	\affiliation{Research Center for Hadron and CSR Physics, Lanzhou University and Institute of Modern Physics of CAS, Lanzhou 730000, China}
	\affiliation{MoE Frontiers Science Center for Rare Isotopes, Lanzhou University, Lanzhou 730000, China}

	\begin{abstract}

As a $D$-wave partner of $\psi(3770)$ identified by the LHCb Collaboration, $\psi_3(3842)$ lies between the $D\bar{D}$ and $D\bar{D^*}$ thresholds. Its non-$D\bar{D}$ decay channels have attracted considerable interest. In this study, we investigate these allowed non-$D\bar{D}$ decays of $\psi_3(3842)$ into $PP$, $PV$, and $VV$ final states using the hadronic loop mechanism, where $P$ and $V$ represent light pseudoscalar and vector mesons, respectively. Our results suggest that these non-$D\bar{D}$ decays of $\psi_3(3842)$ can be significant, with contributions primarily driven by hadronic loops. Notably, the $\rho\pi$ channel stands out as the 
main non-$D\bar{D}$ decay mode, while non-$D\bar{D}$ decay channels involving strange mesons are also sizable. These predictions could be tested in future experiments such as those at LHCb and BESIII.

\end{abstract}

\maketitle
\section{Introduction}

Since the discovery of the first charmonium-like state $X(3872)$ by the Belle Collaboration in 2003 \cite{Belle:2003nnu}, a series of $XYZ$ states have been reported by various experiments, greatly advancing our understanding of hadron spectroscopy \cite{Liu:2013waa, Chen:2016qju, Esposito:2016noz, Chen:2016spr, Guo:2017jvc, Olsen:2017bmm, Liu:2019zoy, Brambilla:2019esw, Chen:2022asf}, particularly in identifying exotic states and constructing conventional hadrons. Moreover, they provide valuable insights into the non-perturbative nature of strong interaction.

{In 2019, the LHCb Collaboration reported a new state, $X(3842)$, observed in the decay channels $X(3842) \to D^0\bar{D}^0$ and $X(3842) \to D^+D^-$, with high statistical significance \cite{LHCb:2019lnr}. The measured resonance parameters are $m_{X(3842)} = 3842.71 \pm 0.16 \pm 0.12\,\text{MeV},\,\Gamma_{X(3842)} = 2.79 \pm 0.51 \pm 0.35\,\text{MeV}$. The predicted mass of the $\psi_3(1^3D_3)$ charmonium state falls within the range of 3806–3912 MeV~\cite{Godfrey:1985xj,Eichten:1980mw,Gupta:1986xt,Fulcher:1991dm,Zeng:1994vj,Ebert:2002pp,Yu:2019kcr,Piemonte:2019cbi,Eichten:2005ga,Radford:2007vd}, and its decay width is expected to lie between 0.8 and 3.0 MeV~\cite{Barnes:2003vb,Barnes:2005pb,Eichten:2005ga}. The measured properties of the $X(3842)$ are therefore consistent with theoretical expectations for the $\psi_3(1^3D_3)$ state, suggesting that $X(3842)$ may be the $D$-wave partner of the $\psi(3770)$ with spin-3, generally denoted as $\psi_3(3842)$}.
Furthermore, the BESIII Collaboration searched for $\psi_3(3842)$ in the process $e^+e^- \to \pi^+\pi^-D^+D^-$ and reported evidence with a statistical significance of $4.2\sigma$~\cite{BESIII:2022quc}.

The non-$D\bar{D}$ decays of $\psi(3770)$ have long posed a puzzle in understanding its decay behavior, with some  experimental and theoretical studies over the past two decades. Since the mass of $\psi(3770)$ is just above the open-charmed threshold of $D\bar{D}$, traditional theories predict that it should primarily decay into pure $D\bar{D}$, with non-$D\bar{D}$ decay channels suppressed by the Okubo-Zweig-Iizuka (OZI) rule. However, experimental data show that the branching ratio for $\psi(3770) \to \text{non}$-$D\bar{D}$ was measured to be $(10.9 \pm 6.9 \pm 9.2)\%$ before 2005 \cite{Rong:2005it}, with several measurements by the BESII Collaboration reporting values of $16.4 \pm 7.3 \pm 4.2\%$ \cite{BES:2006fpf}, $14.5 \pm 1.7 \pm 5.8\%$ \cite{BES:2006dso}, $13.4 \pm 5.0 \pm 3.6\%$ \cite{Ablikim:2007zz}, and $15.1 \pm 5.6 \pm 1.8\%$ \cite{BES:2008vad}. Notably, the CLEO Collaboration reported a measurement of $BR(\psi(3770)\to D\bar{D})=(100.3 \pm 1.4^{+4.8}_{-6.6})\%$ \cite{CLEO:2005mpm}, further adding to the puzzle. These experimental branching ratios for $\psi(3770) \to \text{non-}D\bar{D}$, which range from approximately $10$\% to $15$\%, are significantly larger than theoretical predictions. Meanwhile, the exclusive non-$D\bar{D}$ decay channels of $\psi(3770)$, such as $J/\psi\eta$, $J/\psi\pi\pi$, $\phi\eta$, and $\gamma\chi_{cJ},\,(J=0,\,1)$, are listed by the Particle Data Group (PDG) \cite{ParticleDataGroup:2024cfk}, but their combined branching ratios total less than $2\%$, failing to explain the discrepancy. This suggests that hidden-charm and radiative decays may not be the dominant contributions to the non-$D\bar{D}$ decays of $\psi(3770)$.

Theoretical efforts to understand the puzzle of $\psi(3770)$'s non-$D\bar{D}$ decays have been extensive \cite{Lipkin:1986av,Kuang:1989ub,Ding:1991vu,Achasov:1990gt,Achasov:1991qp,Achasov:1994vh,Achasov:2005qb,Rosner:2001nm,Rosner:2004wy,Voloshin:2005sd,Eichten:2007qx,He:2008xb,Zhang:2009kr,Liu:2009dr,Li:2013zcr,Qi:2025hwd}. In Ref. \cite{Kuang:1989ub}, the authors calculated the hidden-charm decay $\psi(3770) \to J/\psi\pi\pi$ using the QCD multipole expansion, obtaining results consistent with the relevant exclusive measurements \cite{ParticleDataGroup:2024cfk}. Ref. \cite{He:2008xb} explored the decays of $\psi(3770) \to \text{light hadrons}$, and incorporating color-octet contributions, they found $\Gamma(\psi(3770) \to \text{light hadrons}) = 467^{-187}_{+338}$ keV, suggesting a total branching ratio for non-$D\bar{D}$ decays of around $5\%$. Since the dominant decay channel of $\psi(3770)$ is $D\bar{D}$, the hadronic loop mechanism was proposed as a pathway to connect the $\psi(3770)$ state with non-$D\bar{D}$ decay channels \cite{Liu:2009dr,Zhang:2009kr,Li:2013zcr}. The results for $PV$ final states range from $(0.2-1.1)\%$ \cite{Zhang:2009kr,Liu:2009dr} to $(0.04-0.17)\%$ or $(3.38-5.23)\%$ \cite{Li:2013zcr}, depending on the parameters chosen, indicating that the discrepancy between theoretical predictions and experimental data can be significantly alleviated \cite{Liu:2009dr}.

	\begin{figure}[htbp]\centering
		\includegraphics[width=0.48\textwidth]{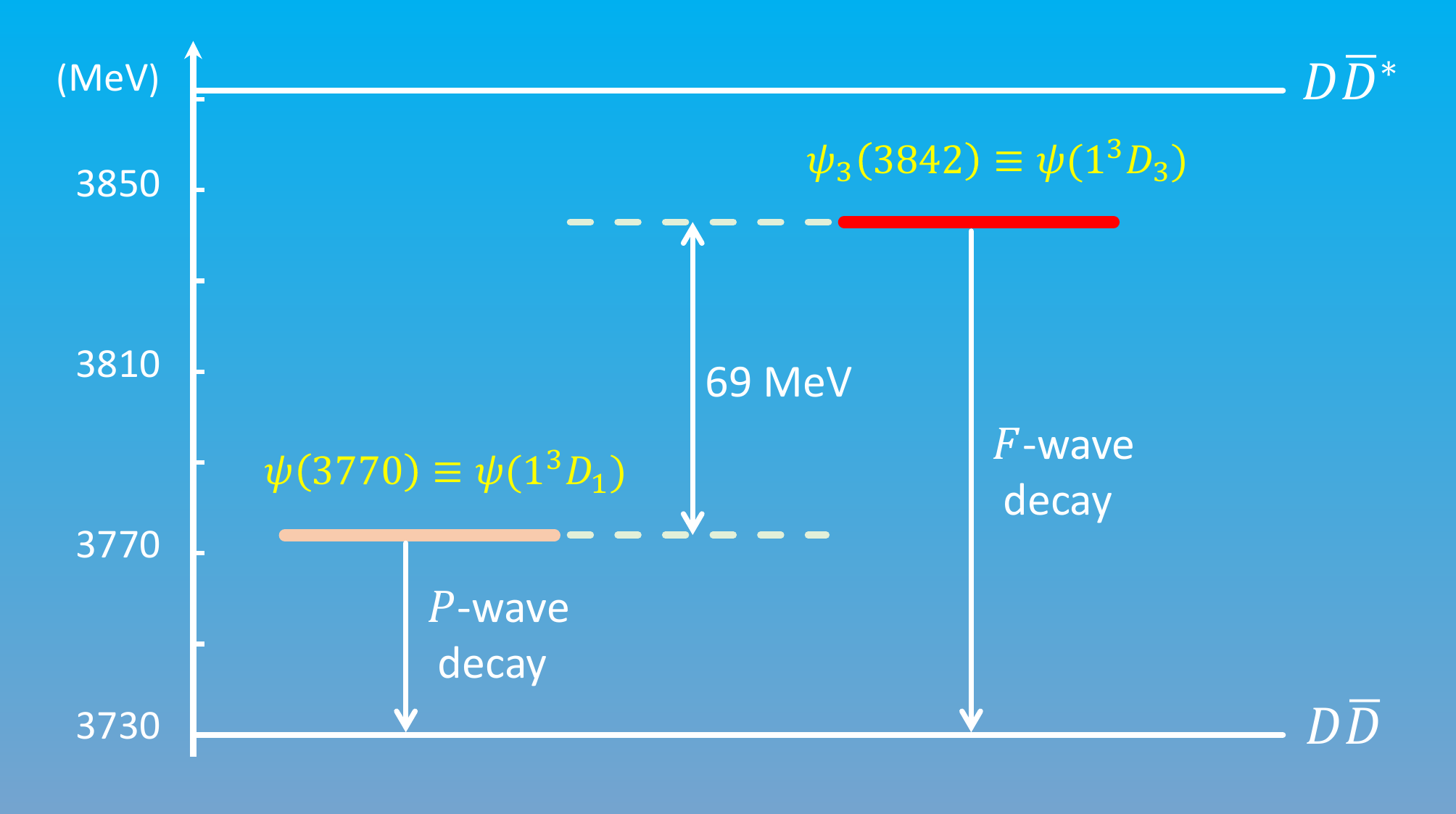}
		\caption{Comparison of the similarities between $\psi(3770)$ and $\psi_3(3842)$, along with their corresponding threshold regions.}
		\label{th}
	\end{figure}

 Notably, there are several similarities between $\psi_3(3842)$ and $\psi(3770)$. Both states share the same spin and orbital angular momentum but differ in their total angular momentum: $\psi_3(3842)$ is a $1^3D_3$ state, while $\psi(3770)$ is a $1^3D_1$ state. The mass gap between them is only 69 MeV, and both lie between the thresholds of $D\bar{D}$ and $D\bar{D^*}$ (see Fig. \ref{th}). Moreover, the only OZI-allowed decay channel for both is $D\bar{D}$, which also serves as their dominant decay mode. The width of $\psi(3770)$ is measured to be $27.5 \pm 0.9$ MeV \cite{ParticleDataGroup:2024cfk}, approximately an order of magnitude larger than that of $\psi_3(3842)$, likely due to the fact that $\psi(3770)$ can decay into $D\bar{D}$ via $P$-wave, while $\psi_3(3842)$ can only decay to $D\bar{D}$ through higher $F$-wave processes.

Given these similarities, along with the unusual behavior of $\psi(3770)$'s non-$D\bar{D}$ decays, it is reasonable to expect that the non-$D\bar{D}$ decays of the $D$-wave charmonium state $\psi_3(3842)$ could also be significant. Theoretical studies in Refs. \cite{Barnes:2003vb,Eichten:2004uh,Barnes:2005pb,Eichten:2005ga,Li:2023cpl,Man:2024mvl} suggest that the non-$D\bar{D}$ decay ratio of $\psi_3(3842)$ could exceed 15\%, raising important questions about the mechanisms of its decay into light meson pairs, which are typical non-$D\bar{D}$ channels, and the magnitude of the branching ratios for these decays.

In this work, we utilize the hadronic loop mechanism to investigate the potential non-$D\bar{D}$ decays of $\psi_3(3842)$ into light meson pairs. The possible final states include three combinations: $PP$, $PV$, and $VV$, where $P$ and $V$ represent pseudoscalar and vector light mesons, respectively. Our results suggest that the non-$D\bar{D}$ decays of $\psi_3(3842)$ into light meson pairs could be substantial, with significant contributions from hadronic loops. We expect that these decay channels may be 
accessible in future experiments.

This paper is organized as follows. After the Introduction, we detail the calculations of non-$D\bar{D}$ decays of $\psi_3(3842)$ into $PP$, $PV$, and $VV$ using the hadronic loop mechanism in Sec. \ref{SecII}. We present our numerical results and discuss their implications in Sec. \ref{SecIII}. Finally, we conclude with a brief summary in Sec. \ref{SecIV}.

	\section{NON-$D\bar D$ DECAYS into light meson pairs OF $\psi_3(3842)$ THROUGH HADRONIC LOOPS}
	\label{SecII}

\begin{figure}[htbp]\centering
		\includegraphics[width=0.48\textwidth]{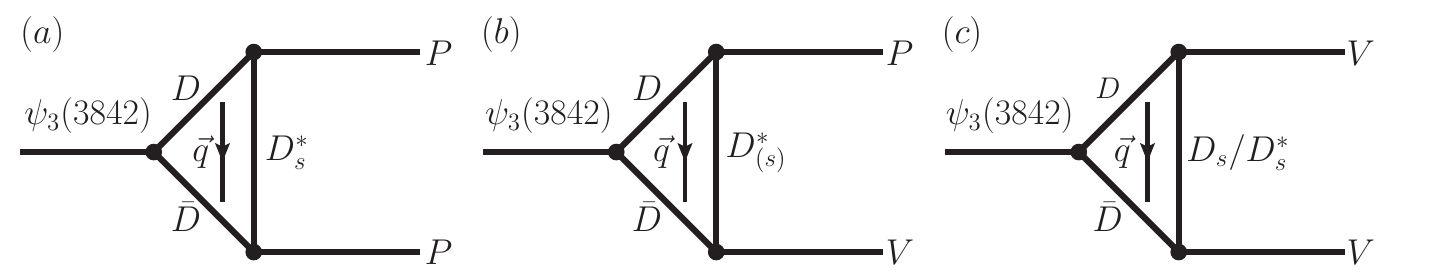}
		\caption{Schematic diagrams illustrating the decays of $\psi_3(3842)$ into $PP$, $PV$, and $VV$ final states via the hadronic loop mechanism.}
		\label{feynman}
	\end{figure}

    One type of non-$D\bar{D}$ decay of $\psi_3(3842)$ arises from its transitions to pairs of light mesons, specifically in the $PP$, $PV$, and $VV$ final states. To quantitatively evaluate these non-$D\bar{D}$ channels, we construct hadronic loops involving charmed mesons, which connect the initial and final states, as shown in Fig. \ref{feynman}. The hadronic loop mechanism, an effective approach for modeling coupled-channel effects \cite{Liu:2006dq,Liu:2009dr,Zhang:2009kr,Li:2013zcr} has been widely applied to study the hadronic decays of higher charmonia and bottomonia \cite{Meng:2007tk,Meng:2008dd,Meng:2008bq,Chen:2011qx,Chen:2011zv,Chen:2011pv,Chen:2011jp,Chen:2014ccr,Wang:2015xsa,Wang:2016qmz,Huang:2017kkg,Zhang:2018eeo,Huang:2018cco,Huang:2018pmk,Li:2021jjt,Bai:2022cfz,Li:2022leg,Bai:2023dhc,Qian:2023taw,Peng:2024xui}. The decay branching ratios predicted by this mechanism are generally in good agreement with experimental measurements.

    In the framework of the hadronic loop mechanism, $\psi_3(3842)$ first decays into a $D\bar{D}$ pair, which subsequently transitions into non-$D\bar{D}$ final states ($PP$, $PV$, and $VV$) through the exchange of a $D_{(s)}^{(*)}$ meson. The decay amplitudes can be expressed as follows:
    \begin{align}
    \label{amp}
    {\mathcal M}=\int\frac{d^{4}q}{(2\pi)^{4}}\frac{{\mathcal V}_{1}{\mathcal V}_{2}{\mathcal V}_{3}}{{\mathcal P}_{1}{\mathcal P}_{2}{\mathcal P}_{E}}\mathcal F^2(q^{2},m_{E}^{2}),
    \end{align}
    where $1/\mathcal{P}_i$ ($i=1,\,2,\,E$) represent the propagators of the intermediate $D_{(s)}^{(*)}$ mesons, and the dipole form factor is given by
    $
    \mathcal{F}(q^2,m_E^2)=\left(\frac{m_E^2-\Lambda^2}{q^2-\Lambda^2}\right)^2$.
    This dipole form factor is adopted to account for the structure and off-shell effects of the exchanged $D_{(s)}^{(*)}$ meson, ensuring the convergence of the amplitude integrals. Here, $q$ and $m_E$ represent the four-momentum and mass of the exchanged $D_{(s)}^{(*)}$ meson, respectively. The cutoff is parametrized as $\Lambda = m_E + \alpha \Lambda_{\text{QCD}}$, where $\Lambda_{\text{QCD}} = 220$ MeV \cite{Liu:2009dr, Zhang:2009kr, Liu:2006dq}, and $\alpha$ is a phenomenological free parameter typically on the order of 1, depending on the specific process \cite{Cheng:2004ru}. The terms $\mathcal{V}_i$ ($i=1,\,2,\,3$) denote the interaction vertices, which are derived using the effective Lagrangian approach. The corresponding Lagrangians are constructed by incorporating heavy quark and chiral symmetries.

The Lagrangian that characterizes the interaction between $\psi_3(1^3D_3)$ and charmed mesons is given by \cite{Li:2021jjt,Casalbuoni:1996pg,Xu:2016kbn}
    \begin{align}
    \label{lagrangian1}
    \mathcal{L}=ig\text{Tr}\Big[D^{(Q\bar Q)\mu\nu}\bar H^{(\bar Qq)}\overset\leftrightarrow\partial_\mu\gamma_\nu\bar H^{(Q\bar q)}\Big]+\text{H.c.}
    \end{align}
    with
    \begin{align}
    D^{(Q\bar Q)\mu\nu}=\frac{1+\slashed{v}}{2}\left(\psi_3^{\mu\nu\alpha}\gamma_\alpha\right)\frac{1-\slashed{v}}{2}.
    \end{align}
   Here, $\psi_3^{\mu\nu\alpha}$ represents the field of $\psi_3(1^3D_3)$, while the field $H$ is constructed from the spin doublet of charmed mesons $(D,\,D^*)$, and is expressed as:
    \begin{align}
    \label{HQq}
    \begin{split}
    H^{Q\bar q}=\frac{1+\slashed v}{2}\Big(D^{*\mu}\gamma_\mu+iD\gamma_5\Big),\\
    H^{\bar Qq}=\Big(\bar D^{*\mu}\gamma_\mu+i\bar D\gamma_5\Big)\frac{1-\slashed v}{2},
    \end{split}
    \end{align}
    and $\bar H^{(Q\bar q)}=\gamma_0 H^{(Q\bar q)\dagger}\gamma_0$, $\bar H^{(\bar Qq)}=\gamma_0 H^{(\bar Qq)\dagger}\gamma_0$.
    Using Eqs. (\ref{lagrangian1})-(\ref{HQq}), we can derive the complete Lagrangian for the $\psi_3$ state interacting with a pair of $D$ mesons, which is given by:
    \begin{align}
    \mathcal{L}_{\psi_3DD}=ig_{\psi_3DD}\psi_3^{\mu\alpha\beta}\Big(\partial_\alpha\partial_\mu D^\dagger\partial_\beta D-\partial_\alpha D^\dagger\partial_\beta\partial_\mu D\Big).
    \end{align}
    
    The Lagrangians governing the interaction of a pair of charmed mesons with a light pseudoscalar ${P}$ and a vector meson ${V}$ are given in Refs. \cite{Li:2021jjt,Casalbuoni:1996pg,Colangelo:2003sa}
    \begin{align}
    \label{LP}
    \mathcal{L}_{{P}}=&ig_P\text{Tr}\Big[H^{(Q\bar q)j}\gamma_\mu\gamma_5\Big(\mathcal{A}^\mu\Big)^i_j\bar H_i^{(Q\bar q)}\Big],\\
    \label{LV}
    \begin{split}
    \mathcal{L}_{{V}}=&i\beta\text{Tr}\Big[H^{(Q\bar q)j}v^\mu\Big(-\rho_\mu\Big) ^i_j\bar H_i^{(Q\bar q)}\Big]\\
    &+i\lambda\text{Tr}\Big[H^{(Q\bar q)j}\sigma^{\mu\nu}F_{\mu\nu}(\rho_j^i)\bar H_i^{(Q\bar q)}\Big],
    \end{split}
    \end{align}
    where $\mathcal{A}^\mu=(\xi^\dagger\partial^\mu\xi-\xi\partial^\mu\xi^\dagger)/2$, with $\xi=e^{i\mathcal{P}/f_\pi}$. $\rho_\mu=ig_V\mathcal{V}_\mu/\sqrt2$, and $F_{\mu\nu}(\rho)=\partial_\mu\rho_\nu-\partial_\nu\rho_\mu+[\rho_\mu,\rho_\nu]$.
  
   The explicit Lagrangians for the $D_{(s)}^{(*)}D_{(s)}^{(*)}P$ and $D_{(s)}^{(*)}D_{(s)}^{(*)}V$ interactions can be derived by expanding Eq. (\ref{LP}) and Eq. (\ref{LV}), respectively, and are given by the following expressions
    \begin{align}
    \begin{split}
    \mathcal{L}_{D^{(*)_{(s)}}D^{(*)}_{(s)}P}=&ig_{DD^*P}\Big(D^{*\dagger}_\mu D-D^\dagger D^*_\mu\Big)\partial^\mu \mathcal{P}\\
    &-g_{D^*D^*P}\varepsilon_{\mu\nu\alpha\beta}D^{*\dagger \nu}\partial^\beta D^{*\alpha}\partial^\mu \mathcal{P}, 
    \end{split}\\
    \begin{split}
    \mathcal{L}_{D_{(s)}^{(*)}D_{(s)}^{(*)}V}=&-ig_{DDV}D^{\dagger}_i\overset{\,\,\leftrightarrow\mu}\partial D^j\Big(\mathcal{V}_\mu\Big)^i_j\\ &-2f_{DD^*V}\varepsilon_{\mu\nu\alpha\beta}\partial^\nu\Big(\mathcal{V}^\beta\Big)^i_j\Big(D^{*\dagger\mu}_i\overset{\,\,\leftrightarrow\alpha}\partial D^j-D^\dagger_i\overset{\,\,\leftrightarrow\alpha}\partial D^{*\mu j}\Big)\\
    &+ig_{D^{*}D^{*}V}D^{*\dagger\nu}_i\overset{\,\,\leftrightarrow\mu}\partial D^{*j}_\nu\Big(\mathcal{V}_\mu\Big)^i_j\\
    &+4if_{D^{*}D^{*}V}D^{*\dagger\mu}_i\Big[\partial_\mu \mathcal{V}^\nu-\partial^\nu \mathcal{V}_\mu\Big]^i_jD^{*j}_\nu,
    \end{split}
    \end{align}
    where $D^{(*)}=\Big(D^{(*)0},\,D^{(*)+},\,D_s^{(*)+}\Big)$ and $D^{(*)\dagger}=\Big(\bar D^{(*)0},\,D^{(*)-},\,D_s^{(*)-}\Big)$.
    The $3\times 3$ matrices for the pseudoscalar octet $\mathcal{P}$ and the vector octet $\mathcal{V}$ are expressed in the following form:
    \begin{align}
        \mathcal{P}=\left(\begin{array}{ccc}\frac{\pi^0}{\sqrt2}+\frac{\cos\theta\,\eta+\sin\theta\,\eta^\prime}{\sqrt6}&\pi^+&K^+\\
        \pi^-&-\frac{\pi^0}{\sqrt2}+\frac{\cos\theta\,\eta+\sin\theta\,\eta^\prime}{\sqrt6}&K^0\\
        K^-&\bar K^0&-\frac{2(\cos\theta\,\eta+\sin\theta\,\eta^\prime)}{\sqrt6}\end{array}\right),
    \end{align}
    
    \begin{align}
        \mathcal{V}=\left(\begin{array}{ccc}\frac{1}{\sqrt2}(\rho^0+\omega)&\rho^+&K^{*+}\\
        \rho^-&\frac{1}{\sqrt2}(-\rho^0+\omega)&K^{*0}\\
        K^{*-}&\bar K^{*0}&\phi\end{array}\right),
    \end{align}
    where $\theta = -19.1^\circ$ denotes the mixing angle between the SU(3) singlet $\eta_1$ and the octet $\eta_8$ \cite{MARK-III:1988crp,DM2:1988bfq}.

We can now derive the interaction vertices $\mathcal{V}_i$ involved in Fig. \ref{feynman} using the Lagrangians discussed above, which are given by
    \begin{align}
    \begin{split}
		&\langle D(q_1)\bar D(q_2)|\psi_3(p)\rangle=g_{\psi_3DD}\epsilon_{\psi_3}^{\mu\alpha\beta}(q_{2\mu}-q_{1\mu})q_{1\alpha}q_{2\beta},\\
		&\langle P(p_1)D^*_{(s)}(q)|D(q_1)\rangle=-g_{DD_{(s)}^*P}\epsilon_{D_{(s)}^*}^{*\mu} p_{1\mu},\\
		&\langle P(p_2)|D^*_{(s)}(q)\bar D(q_2)\rangle=g_{DD_{(s)}^*P}\epsilon_{D_{(s)}^*}^\mu p_{2\mu},\\
		&\langle V(p_1)D_{(s)}(q)|D(q_1)\rangle=-g_{DD_{(s)}V}\epsilon_{V}^{*\mu}(q_{\mu}+q_{1\mu}),\\
		&\langle V(p_2)|D_{(s)}(q)\bar D(q_2)\rangle=-g_{DD_{(s)}V}\epsilon_{V}^{*\mu}(q_{\mu}-q_{2\mu}),\\
		&\langle V(p_1)D^*_{(s)}(q)|D(q_1)\rangle=-2f_{DD_{(s)}^*V}\varepsilon_{\mu\nu\alpha\beta}\epsilon_{V}^{*\beta}\epsilon_{D^*_{(s)}}^{*\mu}p_1^\nu(q^{\alpha}+q_1^\alpha),\\
		&\langle V(p_2)|D^*_{(s)}(q)\bar D(q_2)\rangle=2f_{DD_{(s)}^*V}\varepsilon_{\mu\nu\alpha\beta}\epsilon_{V}^{*\beta}\epsilon_{D^*_{(s)}}^{\mu}p_2^\nu(q^{\alpha}-q_2^\alpha).\\
        \end{split}
	\end{align}
	
	The decay amplitudes for $\psi_3(3842)$ into $PP$, $PV$, and $VV$ via the hadronic loop mechanism, as shown in Fig. \ref{feynman}, can be expressed as
	 \begin{align}
    \mathcal{M}^{(a)}_{PP}=&i^3\int\frac{d^4q}{(2\pi)^4}\Big[g_{\psi_3DD}\epsilon_{\psi_3}^{\mu\alpha\beta}(q_{2\mu}-q_{1\mu})q_{1\alpha}q_{2\beta}\Big]\\
    &\times(-g_{DD_s^*P}p_{1\nu})(g_{DD_s^*P}p_{2\lambda})\frac1{q_1^2-m_D^2}\nonumber\\
    &\times\frac1{q_2^2-m_D^2}\frac{-g^{\nu\lambda}+q^\nu q^\lambda/m_{D_s^*}^2}{q^2-m_{D_s^*}^2}\mathcal F^2(q^{2},m_{D_s^*}^{2}),\nonumber
    \end{align}

    \begin{align}
    \mathcal{M}^{(b)}_{PV}=&i^3\int\frac{d^4q}{(2\pi)^4}\Big[g_{\psi_3DD}\epsilon_{\psi_3}^{\mu\alpha\beta}(q_{2\mu}-q_{1\mu})\\
	&\times q_{1\alpha}q_{2\beta}\Big](-g_{DD_{(s)}^*P}p_{1\tau})\nonumber\\
    &\times\Big[2f_{DD_{(s)}^*V}\varepsilon_{\nu\lambda\theta\delta}p_2^\lambda(q^\theta-q_2^\theta)\epsilon_V^{*\delta}\Big]\nonumber\\
    &\times\frac1{q_1^2-m_D^2}\frac1{q_2^2-m_D^2}\frac{-g^{\tau\nu}+q^\tau q^\nu/m_{D^*}^2}{q^2-m_{D_{(s)}^*}^2}\nonumber\\
    &\times\mathcal F^2(q^{2},m_{D_{(s)}^*}^{2}),\nonumber
    \end{align}

    \begin{align}
    \mathcal{M}^{(c)}_{VV}=&\mathcal{M}^{(c_1)}_{VV}+\mathcal{M}^{(c_2)}_{VV}\\
    =&i^3\int\frac{d^4q}{(2\pi)^4}\Big[g_{\psi_3DD}\epsilon_{\psi_3}^{\mu\alpha\beta}(q_{2\mu}-q_{1\mu})q_{1\alpha}q_{2\beta}\Big]\nonumber\\
    &\times\Big[-g_{DD_sV}(q_{\xi}+q_{1\xi})\epsilon_{V}^{*\xi}\Big]\nonumber\\
    &\times\Big[-g_{DD_sV}(q_{\delta}-q_{2\delta})\epsilon_{V}^{*\delta}\Big]\nonumber\\
    &\times\frac1{q_1^2-m_D^2}\frac1{q_2^2-m_D^2}\frac1{q^2-m_{D_s}^2}\nonumber\\
    &\times\mathcal F^2(q^{2},m_{D_s}^{2})\nonumber\\
    &+i^3\int\frac{d^4q}{(2\pi)^4}\Big[g_{\psi_3DD}\epsilon_{\psi_3}^{\mu\alpha\beta}(q_{2\mu}-q_{1\mu})q_{1\alpha}q_{2\beta}\Big]\nonumber\\
    &\times\Big[-2f_{DD_s^*V}\varepsilon_{\theta\tau\nu\xi}p_1^\tau(q^{\nu}+q^{\nu}_{1})\epsilon_{V}^{*\xi}\Big]\nonumber\\
    &\times\Big[2f_{DD_s^*V}\varepsilon_{\iota\lambda\zeta\delta}p_2^\lambda(q^{\zeta}-q^{\zeta}_{2})\epsilon_{V}^{*\delta}\Big]\nonumber\\
    &\times\frac1{q_1^2-m_D^2}\frac1{q_2^2-m_D^2}\frac{-g^{\theta\iota}+q^\theta q^\iota/m_{D_s^*}^2}{q^2-m_{D_s^*}^2}\nonumber\\
    &\times\mathcal F^2(q^{2},m_{D_s^*}^{2}),\nonumber
    \end{align}
    where $\mathcal{M}^{(c_1)}_{VV}$ and $\mathcal{M}^{(c_2)}_{VV}$ represent the amplitudes corresponding to the $D_s$ exchange and $D_s^*$ exchange, respectively, in the decay process $\psi_3(3842) \to VV$, as illustrated in Fig. \ref{feynman}(c). The non-$D\bar{D}$ decay widths of $\psi_3(3842)$ can then be calculated using the following formula
    \begin{align}
		\Gamma=\frac17\frac{|\vec{p}_1|}{8\pi m_{\psi_3(3842)}^2}\sum_{spin}\Big|{\mathcal{M}}^{\text{Total}}\Big|^2,
	\end{align}
	where $\vec{p}_1$ is the three-momentum of the final state in the center-of-mass frame of the initial state. The factor of 1/7 accounts for the averaging over the polarizations of the initial state, while the sum $\sum_{\text{spin}}$ denotes the summation over the polarizations of the final state.
	

    \section{NUMERICAL RESULTS}\label{SecIII}
In this section, we will explain how the relevant coupling constants are determined and present the numerical results for the non-$D\bar{D}$ decays of $\psi_3(3842)$ into light meson pairs of $PP$, $PV$, and $VV$. We note that the $PP$ and $VV$ final states are limited to strange meson pairs, specifically $K\bar{K}$ and $K^*\bar{K}^*$, respectively, while other light unflavored meson pairs are suppressed by C-parity. The $PV$ final states include $\omega\eta^{(\prime)}$, $\rho\pi$, and $K\bar{K}^*$. We provide a detailed list of the intermediate hadronic loops involving charmed mesons for these final states in Table \ref{loops}, with the corresponding Feynman diagrams shown in Fig. \ref{feynman}. The branching ratios for the decays of $\psi_3(3842)$ into each neutral and charged component of the same final state are considered equal, assuming isospin symmetry.
    
    \begin{table}[htbp]
		\centering
		\caption{The detailed intermediate loops connecting the initial state $\psi_3(3842)$ to the final states of $PP$, $PV$, and $VV$ as depicted in Fig. \ref{feynman}.}
		\label{loops}
		\renewcommand\arraystretch{1.3}
		\setlength{\tabcolsep}{8pt}
		\setlength{\arrayrulewidth}{0.5pt}
		\begin{tabular}{c|c|c|c}
			\toprule[1pt]
			\toprule[0.5pt]
			\multicolumn{3}{c|}{Final states} &Hadronic loops\\
			\midrule[0.5pt]
			\multirow{2}{*}{$PP$}  &\multirow{2}{*}{$K\bar K$} &$\bar K^0(p_1)K^0(p_2)$ &$D^+(q_1)D^-(q_2)D_s^{*+}(q)$\\
			\cline{3-4}
			&{} &$K^-(p_1)K^+(p_2)$ &$D^0(q_1)\bar D^0(q_2)D_s^{*+}(q)$\\
			\midrule[0.5pt]
			\multirow{16}{*}{$PV$} &\multirow{4}{*}{$\omega\eta^{(\prime)}$} &\multirow{4}{*}{$\eta^{(\prime)}(p_1)\omega(p_2)$} &$D^0(q_1)\bar D^0(q_2)D^{*0}(q)$\\
			\cline{4-4}
			&{} &{} &$\bar D^0(q_1)D^0(q_2)\bar D^{*0}(q)$\\
			\cline{4-4}
			&{} &{} &$D^+(q_1)D^-(q_2)D^{*+}(q)$\\
			\cline{4-4}
			&{} &{} &$D^-(q_1)D^+(q_2)D^{*-}(q)$\\
			\cline{2-4}
			&\multirow{8}{*}{$\rho\pi$} &\multirow{4}{*}{$\pi^0(p_1)\rho^0(p_2)$} &$D^0(q_1)\bar D^0(q_2)D^{*0}(q)$\\
			\cline{4-4}
			&{} &{} &$\bar D^0(q_1)D^0(q_2)\bar D^{*0}(q)$\\
			\cline{4-4}
			&{} &{} &$D^+(q_1)D^-(q_2)D^{*+}(q)$\\
			\cline{4-4}
			&{} &{} &$D^-(q_1)D^+(q_2)D^{*-}(q)$\\
			\cline{3-4}
			&{} &\multirow{2}{*}{$\pi^-(p_1)\rho^+(p_2)$} &$D^-(q_1)D^+(q_2)\bar D^{*0}(q)$\\
			\cline{4-4}
			&{} &{} &$D^0(q_1)\bar D^0(q_2)D^{*+}(q)$\\
			\cline{3-4}
			&{} &\multirow{2}{*}{$\pi^+(p_1)\rho^-(p_2)$} &$D^+(q_1)D^-(q_2)D^{*0}(q)$\\
			\cline{4-4}
			&{} &{} &$\bar D^0(q_1)D^0(q_2)\bar D^{*-}(q)$\\
			\cline{2-4}
			&\multirow{4}{*}{$K\bar K^*$} &$K^0(p_1)\bar K^{*0}(p_2)$ &$D^-(q_1)D^+(q_2)D_s^{*-}(q)$\\
			\cline{3-4}
			&{} &$\bar K^0(p_1)K^{*0}(p_2)$ &$D^+(q_1)D^-(q_2)D_s^{*+}(q)$\\
			\cline{3-4}
			&{} &$K^+(p_1)K^{*-}(p_2)$ &$\bar D^0(q_1)D^0(q_2)D_s^{*-}(q)$\\
			\cline{3-4}
			&{} &$K^-(p_1)K^{*+}(p_2)$ &$D^0(q_1)\bar D^0(q_2)D_s^{*+}(q)$\\
			\midrule[0.5pt]
			\multirow{4}{*}{$VV$}  &\multirow{4}{*}{$K^*\bar K^*$} &\multirow{2}{*}{$\bar K^{*0}(p_1)K^{*0}(p_2)$} &$D^+(q_1)D^-(q_2)D_s^{+}(q)$\\
			\cline{4-4}
			&{} &{} &$D^+(q_1)D^-(q_2)D_s^{*+}(q)$\\
			\cline{3-4}
			&{} &\multirow{2}{*}{$K^{*-}(p_1)K^{*+}(p_2)$} &$D^0(q_1)\bar D^0(q_2)D_s^{+}(q)$\\
			\cline{4-4}
			&{} &{} &$D^0(q_1)\bar D^0(q_2)D_s^{*+}(q)$\\
			\bottomrule[0.5pt]
			\bottomrule[1pt]
		\end{tabular}
	\end{table}	

    The coupling constant $g_{\psi_3DD}$ is determined by matching the theoretical decay width $\Gamma(\psi_3(3842) \to D\bar{D}) = 2.35 \, \text{MeV}$ \cite{Li:2023cpl}, the resulting value is $g_{\psi_3DD} = 20.52 \, \text{GeV}^{-2}$. The coupling constants $g_{DD_{(s)}^*P}$ and $g_{DD_{(s)}^*V}$ are related to the global coupling constant $g_P$ and the parameters $\beta$ and $\lambda$, which arise from expanding the Lagrangians in Eq. (\ref{LP}) and Eq. (\ref{LV}), respectively. Specifically, we have
    \begin{align}
    \begin{split}
    g_{DD^*\pi^0}&=\frac{1}{\sqrt2}g_{DD^*\pi^\pm}=\frac{g_P\sqrt{2m_Dm_{D^*}}}{f_\pi},\\
    g_{DD^*\eta}&=\frac{\cos\theta}{\sqrt6}\frac{2g_P\sqrt{m_Dm_{D^*}}}{f_\pi},\\
    g_{DD^*\eta^\prime}&=\frac{\sin\theta}{\sqrt6}\frac{2g_P\sqrt{m_Dm_{D^*}}}{f_\pi},\\
    g_{DD_s^*K}&=\frac{2g_P\sqrt{m_Dm_{D_s^*}}}{f_\pi},
    \end{split}
    \end{align}

     \begin{align}
     \begin{split}
        g_{DD\omega}&=g_{DD\rho^0}=\frac{1}{\sqrt2}g_{DD\rho^\pm}=\frac{\beta g_V}{2},\\
        g_{DD_sK^*}&=\sqrt{\frac{m_{D_s}}{m_D}}\frac{\beta g_V}{\sqrt2},\\
        f_{DD^*\omega}&=f_{DD^*\rho^0}=\frac{1}{\sqrt2}f_{DD^*\rho^\pm}=\frac{\lambda g_V}{2},\\
        f_{DD_s^*K^*}&=\sqrt{\frac{m_{D_s^*}}{m_{D^*}}}\frac{\lambda g_V}{\sqrt2},
        \end{split}
    \end{align}
   where $g_P = 0.569$ is determined by fitting the experimentally measured partial width of $\Gamma(D^{*+} \to D^0 \pi^+)$ \cite{ParticleDataGroup:2024cfk}, $\beta = 0.9$, $\lambda = 0.56 \, \text{GeV}^{-1}$, and $g_V = m_\rho / f_\pi$, with $f_\pi = 132$ MeV \cite{Cheng:1992xi,Yan:1992gz,Wise:1992hn,Burdman:1992gh}.

{Besides the coupling constants mentioned above, there remains one undetermined parameter, $\alpha$, which is introduced in the form factor. Recently, the BESIII Collaboration reported the first observation of the decay process $\psi(3770) \to K_S^0 K_L^0$, with a measured branching ratio of $(2.63^{+1.40}_{-1.59}) \times 10^{-5}$ \cite{BESIII:2023zsk}. This decay can proceed via $\psi(3770) \to D^+ D^- \to K^0 \bar{K}^0$, mediated by exchange of a $D_s^{*+}$ meson within the same hadronic loop framework. By reproducing the observed branching ratio of $\psi(3770) \to K_S^0 K_L^0$, we can constrain the allowed range of the parameter $\alpha$.

The interaction vertex for $\psi(3770)D\bar D$ is given by \cite{Liu:2009dr,Zhang:2009kr,Li:2013zcr}
\begin{align}
\langle D^+(q_1) D^-(q_2)|\psi(3770)(p)\rangle=g_{\psi(3770)DD}\epsilon_{\psi(3770))}^{\mu}(q_{2\mu}-q_{1\mu}),
\end{align}
where the coupling constant $g_{\psi(3770)DD}=11.96$ is determined based on the average branching ratio of $\psi(3770)\to D\bar D$ listed by PDG.

The value of the parameter $\alpha$ is constrained to the range 1.0--1.3 by matching the measured branching ratio of $\psi(3770) \to K_S^0 K_L^0$. Given the similarities between $\psi(3770)$ and $\psi_3(3842)$, as discussed in the Introduction, we adopt the same $\alpha$ range to predict the non-$D\bar{D}$ decays of $\psi_3(3842)$ into light meson pairs.}

\begin{figure}[htbp]\centering
		\includegraphics[width=0.47\textwidth]{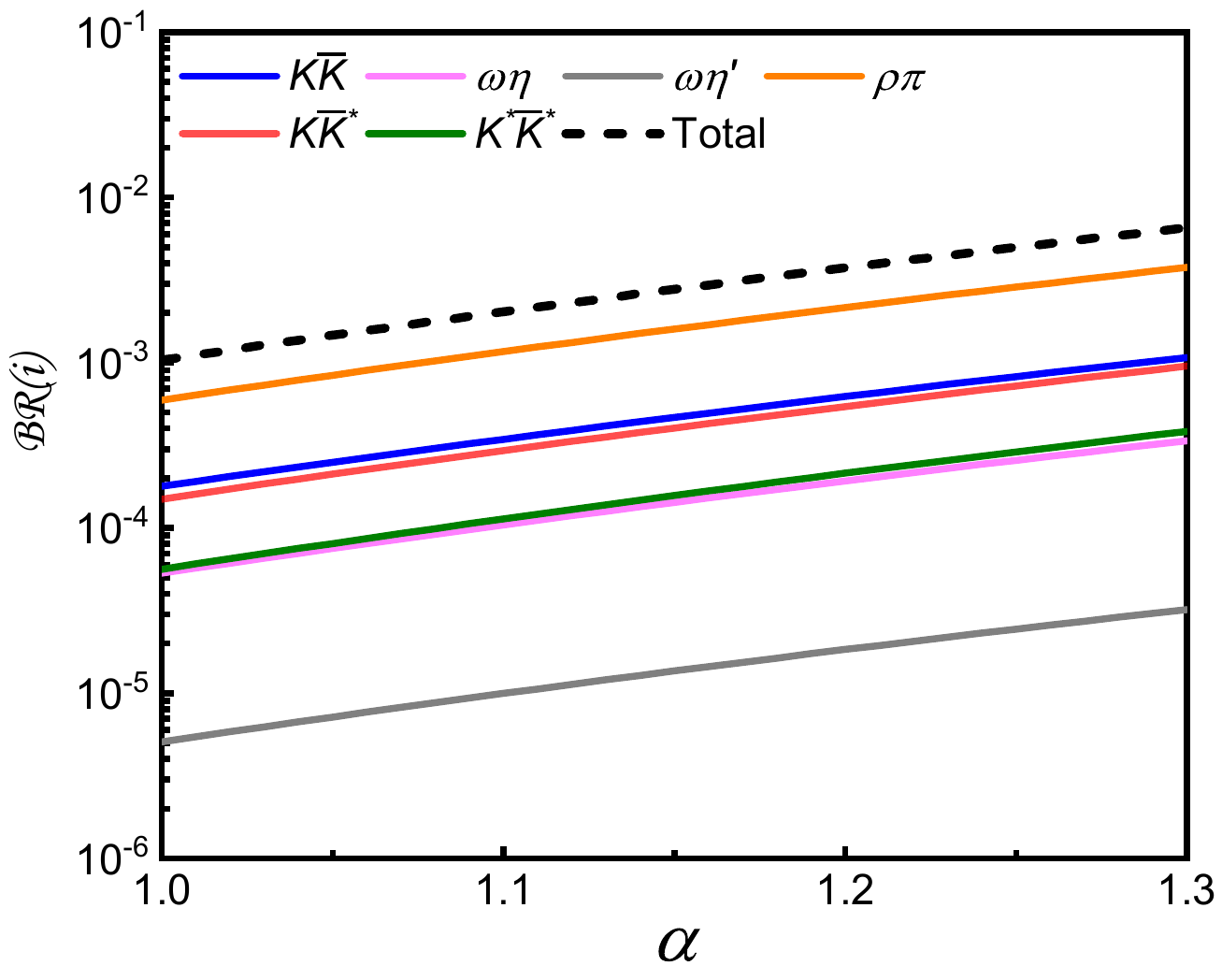}
		\caption{The $\alpha$ parameter dependence of the branching ratios for the non-$D\bar D$ decays of $\psi_3(3842)$ into $PP$, $PV$, and $VV$, with $i$ representing the different final states.}
		\label{figbr}
	\end{figure}

With above preparations, we calculate the numerical branching ratios of $\psi_3(3842)$ decays into $PP$, $PV$, and $VV$ final states. {The results are presented in Fig. \ref{figbr}, with the $\alpha$ parameter varying from 1.0 to 1.3.
We find that the total non-$D\bar D$ branching ratio is in the range of $(0.1 - 0.7)\%$.} The result suggest that $\psi_3(3842)$ exhibits non-$D\bar D$ decay behavior similar to those of $\psi(3770)$ \cite{Liu:2009dr,Zhang:2009kr,Li:2013zcr}, implying the potential for sizable non-$D\bar D$ decays in $\psi_3(3842)$. Among the decay modes of $\psi_3(3842)$, the $\rho \pi$ channel is the largest, primarily due to the sufficient phase space, relatively large coupling constants, and three different charge components, as shown in Table \ref{loops}. 

     \begin{figure}[htbp]\centering
		\includegraphics[width=0.47\textwidth]{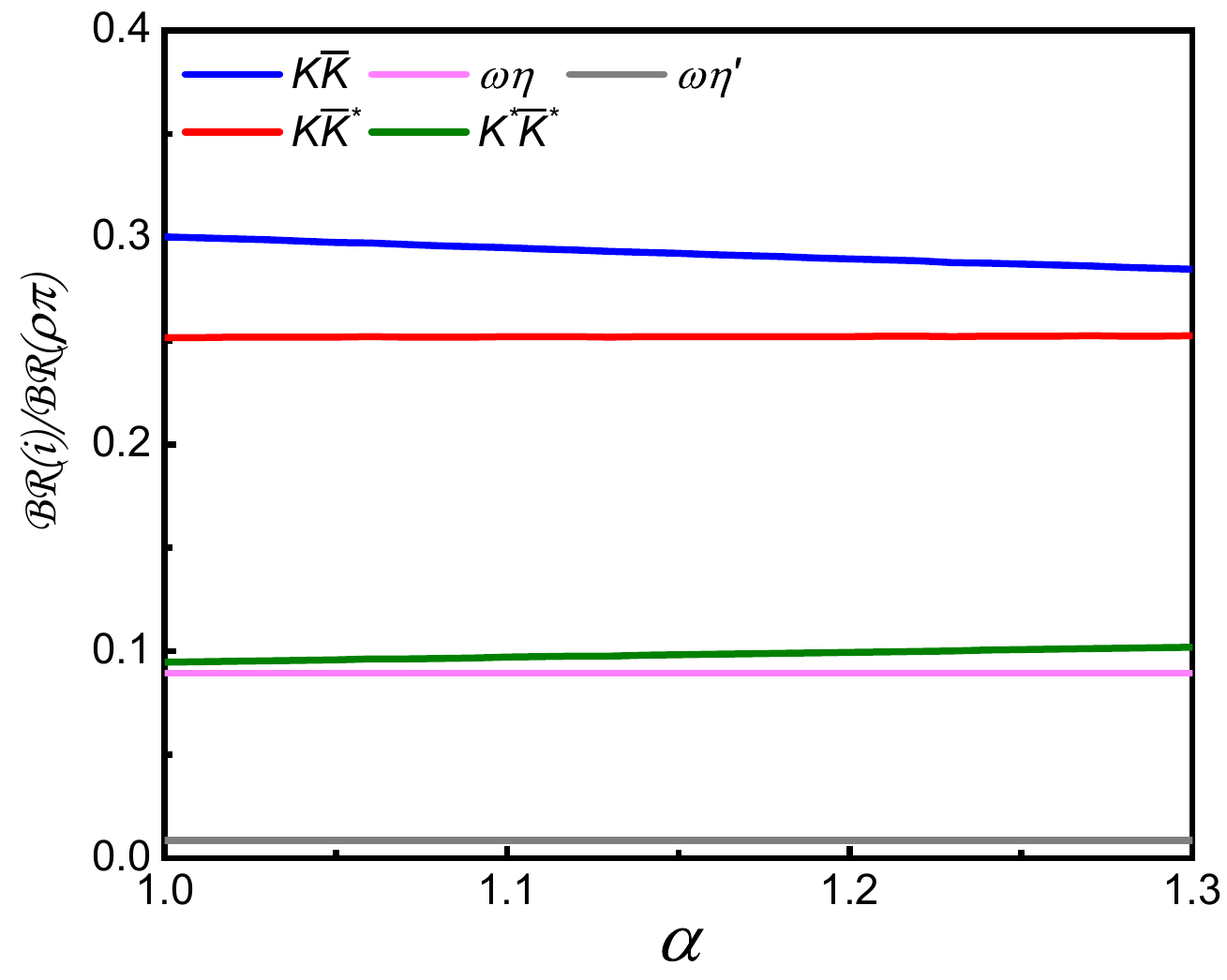}
		\caption{The $\alpha$ parameter dependence of the ratios $\mathcal{BR}(i)/\mathcal{BR}(\rho\pi)$. Here, $i$ represent the different final states of $PP$, $PV$, and $VV$.}
		\label{figbrr}
	\end{figure}

     The $\rho\pi$ channel as the dominant non-$D\bar D$ decay mode of $\psi_3(3842)$, is expected to be observed in future experiments. As shown in Eq. (\ref{RF}) and Fig. \ref{figbrr}, we also present the relative branching fractions of the other decay processes compared to that of $\psi_3(3842) \to \rho\pi$, {with $\alpha$ parameter in the range of 1.0--1.3.

    \begin{align}\label{RF}
		\mathcal{BR}(K\bar K)/\mathcal{BR}(\rho\pi)&=(2.8-3.0)\times10^{-1},\nonumber\\
            \mathcal{BR}(\omega\eta)/\mathcal{BR}(\rho\pi)&=8.9\times10^{-2},\nonumber\\
		\mathcal{BR}(\omega\eta^\prime)/\mathcal{BR}(\rho\pi)&=(8.5-8.6)\times10^{-3},\\ 
		\mathcal{BR}(K\bar K^*)/\mathcal{BR}(\rho\pi)&=2.5\times10^{-1},\nonumber\\
		\mathcal{BR}(K^*\bar K^*)/\mathcal{BR}(\rho\pi)&=(9.5-10.2)\times10^{-2}.\nonumber
	\end{align}}

	 The relative branching fractions exhibit only weak dependence on the $\alpha$ parameter. Channels involving strange mesons are also noteworthy, with the branching fraction ratios approximately given by {$\mathcal{BR}(K\bar K):\mathcal{BR}(K\bar K^*):\mathcal{BR}(K^*\bar K^*) \approx 2.9:2.5:1.0$}. Contributions from $\omega\eta^{(\prime)}$ are negligible compared to the dominant $\rho\pi$ channel, with the ratio between them found to be $\mathcal{BR}(\omega\eta):\mathcal{BR}(\omega\eta^\prime) \approx 10:1$ in our calculations. These results may assist future experimental efforts, such as those conducted by the LHCb and BESIII Collaborations, in further confirming the $\psi_3(3842)$ and investigating its non-$D\bar D$ decay modes.

    \section{SUMMARY}\label{SecIV}

   For charmonia above the $D\bar{D}$ thresholds, the dominant decay channels are typically the OZI-allowed open-charm decays, such as $D\bar{D}$, while other decay modes are expected to be negligible. However, experimental measurements suggest that the branching ratio for non-$D\bar{D}$ decays of $\psi(3770)$ can reach $(10-15)\%$, which is significantly larger than theoretical predictions. This long-standing puzzle, deeply linked to the non-perturbative effects of the strong interaction, has drawn considerable attention from both theorists and experimentalists seeking to understand the exotic behaviors of the non-$D\bar{D}$ decays of $\psi(3770)$. It is worth noting that in Refs. \cite{Zhang:2009kr,Liu:2009dr,Li:2013zcr}, the introduction of the hadronic loop mechanism significantly reduces the discrepancy between theoretical predictions and experimental data for the non-$D\bar{D}$ decays of $\psi(3770)$. Furthermore, the hadronic loop mechanism, which describes the transition from $D\bar{D}$ to non-$D\bar{D}$ channels, offers a natural explanation for the puzzling non-$D\bar{D}$ decays of $\psi(3770)$ and plays a crucial role in studying other charmonia above the thresholds of charmed meson pairs.

In this work, we focus on the $D$-wave charmonium $\psi_3(3842)$, a spin-3 partner of $\psi(3770)$ as identified by the LHCb collaboration \cite{LHCb:2019lnr}. It lies in an energy region between the thresholds of $D\bar{D}$ and $D\bar{D^*}$, suggesting that its dominant decay mode is $D\bar{D}$. By comparing the similarities between $\psi_3(3842)$ and $\psi(3770)$, we conjecture that the non-$D\bar{D}$ decays of $\psi_3(3842)$ may also be significant, and investigate them using the hadronic loop mechanism. With the corresponding effective Lagrangians, we systematically calculate the potential non-$D\bar{D}$ decays of $\psi_3(3842)$ into $PP$, $PV$, and $VV$ final states. Our results suggest that the non-$D\bar{D}$ decays of $\psi_3(3842)$ are sizable, with significant contributions from the hadronic loops. Among these, the $\rho\pi$ channel is the dominant non-$D\bar{D}$ decay mode, while channels involving strange mesons are also noteworthy. We are optimistic that our findings will help future experiments, such as those conducted by the LHCb and BESIII collaborations, to further confirm $\psi_3(3842)$ and explore its non-$D\bar{D}$ decays.

	\section*{ACKNOWLEDGMENTS}
	
	This work is supported by the National Natural Science Foundation of China under Grant Nos. 12335001 and 12247101,  the ‘111 Center’ under Grant No. B20063, the Natural Science Foundation of Gansu Province (No. 22JR5RA389,  No.25JRRA799), the fundamental Research Funds for the Central Universities, and the project for top-notch innovative talents of Gansu province.

	\vfil

\end{document}